\documentclass[aps,prx,superscriptaddress,floatfix,longbibliography]{revtex4-2}

\usepackage[utf8]{inputenc}
\usepackage[T1]{fontenc}
\usepackage{microtype}
\usepackage{amsmath}
\usepackage{amsfonts}
\usepackage{amssymb}
\usepackage{mathtools}
\usepackage{graphicx}
\usepackage{bm}
\usepackage{titlesec}
\usepackage{siunitx}
\usepackage{setspace}
\usepackage{nowidow}

\usepackage{xcolor}
\usepackage[colorlinks,
    citecolor=blue,
    linkcolor=blue,
    urlcolor=blue]{hyperref}

\newcommand{\figref}[2]{\hyperref[#1]{Fig.~\ref{#1}#2}}
\newcommand{\secref}[1]{\hyperref[#1]{Sec.~\ref{#1}}}
\newcommand{\eref}[1]{\hyperref[#1]{Eq.~\ref{#1}}}
\newcommand{\aref}[1]{\hyperref[#1]{Appendix~\ref{#1}}}
\setcitestyle{authoryear,round}

\begin{document}

\titleformat{\section}{\raggedright\bfseries}{\thesection.}{1em}{}` 

\title{Active filtering: a predictive function of recurrent circuits of sensory cortex}

\author{Mark H. Histed}
\email{mark.histed@nih.gov}
\affiliation{NIMH Intramural Program, National Institutes of Health, Bethesda MD 20892}

\date{\today: Preprint}

\thanks{I am grateful to Tyler Sloan (Quorumetrix Studio, \url{https://www.quorumetrix.com/}) for the visualization in Fig.~\ref{Fig:1}C, Ciana Deveau for work on figures, and Ms.~Deveau, Haleigh Mulholland, Jonathan O'Rawe and the rest of the Histed lab for comments and discussion. This work was funded by the National Institutes of Health NIMH Intramural Program (ZIAMH002956).}

\begin{abstract}
Our brains encode many features of the sensory world into memories: we can sing along with songs we have heard before, interpret spoken and written language composed of words we have learned, and recognize faces and objects. Where are these memories stored? Each area of the cerebral cortex has a huge number of local, recurrent, excitatory--excitatory synapses, as many as 500 million per cubic millimeter. Here I review evidence that cortical recurrent connectivity in sensory cortex is a substrate for sensory memories. Evidence suggests that the local recurrent network encodes the structure of natural sensory input, and that it does so via active filtering, transforming network inputs to boost or select those associated with natural sensation. This is a form of predictive processing, in which the cortical recurrent network selectively amplifies some input patterns and attenuates others, and a form of memory.
\end{abstract}

\maketitle

\section{Active filtering: a predictive computation performed by recurrent networks}

\begin{figure}[!ht]\centering
    \includegraphics[width=0.6\textwidth]{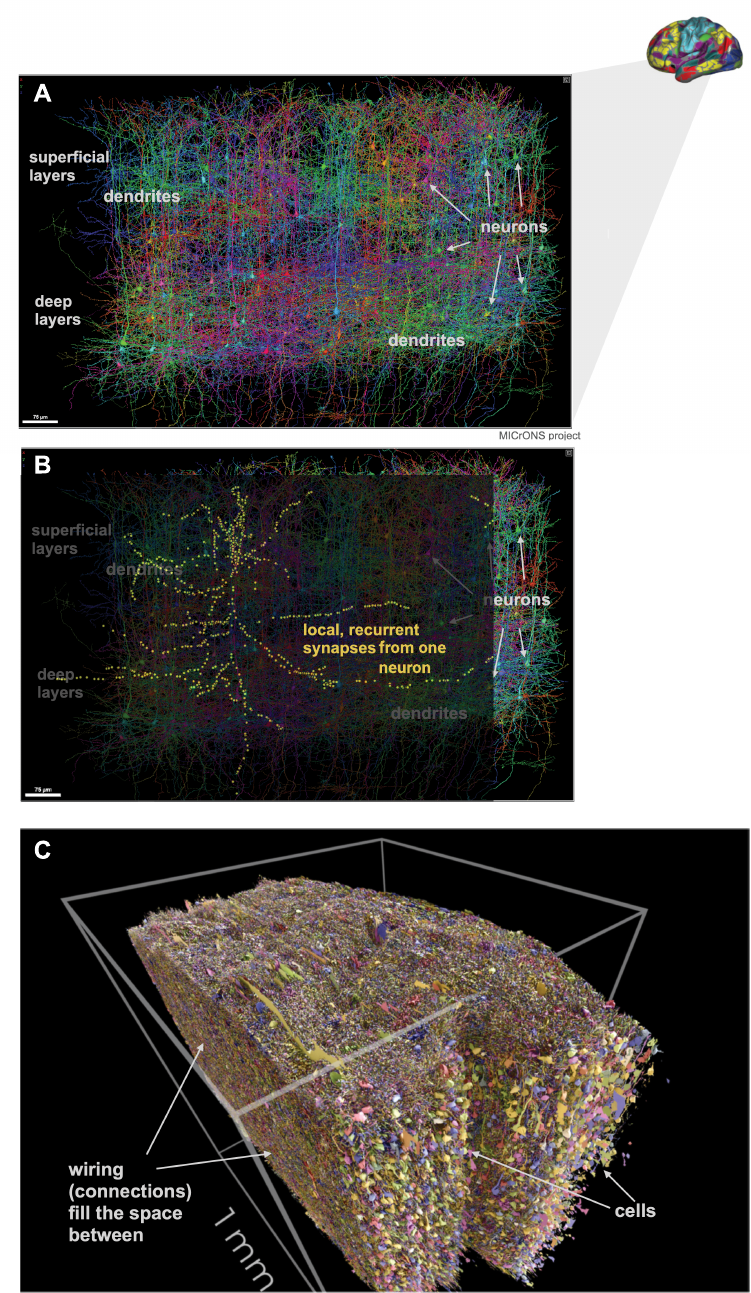}
    \caption{Local recurrent excitatory--excitatory connections in the cortex are dense and numerous.
        \textbf{A}, side view of a sparse set of neurons constructed via serial-section electron microscopy \citep{MICrONS}, see also \url{https://microns-explorer.org}. Each cell is shown in a different color. A large amount of dendritic processes from multiple cells can be seen. Some cell bodies are highlighted by arrows (white text: ``neurons''). Data from the mouse cortex, but recurrent structure is broadly similar in the human (schematic, top right).
        \textbf{B}, Same data with synapses formed by just one excitatory cell superimposed. Each yellow point is a postsynaptic density reconstructed by the MICrONS project by following the cells' axon. Source cell is in layer 2/3. The fact that just one excitatory cell forms hundreds of local connections implies that much of the cortical neuropil, the volume between cell bodies in the cortex, is dedicated to local, recurrent connectivity.
        \textbf{C}, Another view of the same data, this time visualized with the entire volume shown, not just a sparse set of neurons as in A, B. Towards bottom right, neural cell bodies have been emphasized by removing some of the neuropil around them for visualization purposes. The cortical neuropil is filled with ``wires,'' dendrites, and axons, and much of this wiring is dedicated to local connectivity.}\label{Fig:1}
\end{figure}

The large majority of neurons in the cerebral cortex are excitatory neurons \citep{Tremblay2016-he,Wonders2006-hj}. These excitatory neurons are densely connected locally (Fig.~\ref{Fig:1}), with each cortical excitatory cells receiving thousands of input synapses, and an estimated 50--80\% of those synapses arising from other excitatory neurons less than a millimeter away \citep{Braitenberg2013-ix}. What is the purpose of these local recurrent connections?

The anatomy of recurrent connectivity hints broadly that recurrence has some significant role in cortical function. First, dense recurrent connectivity is common across cortical areas --- excitatory--excitatory recurrence is a feature of all cerebral cortical regions \citep{Elston2003-xd}. And creating and maintaining these synaptic connections is metabolically expensive \citep{Li2022-dq}.

This review puts forth a proposal for the role of cortical recurrent connections in the function of the cortex, focusing on sensory cortex. Several pieces of evidence suggest that different sensory cortices, at least visual and somatosensory cortex, actively filter inputs --- that is, they selectively amplify certain patterns of input and suppress others. This is a predictive function, where input patterns that are associated with normal or natural inputs are boosted, and others are suppressed. And it is a form of memory, where the network encodes the features of the sensory world and uses that stored information to boost inputs that are statistically similar to features of the natural world.

A key aspect of this active filtering computation is that it operates on patterns of input, not just on input to single neurons. During normal cortical operation, long-range inputs (external to the local recurrent network) arrive to a population of neurons. For example, in the visual system, even a small spot of light activates several retinal ganglion cells, and changes the activity of multiple lateral geniculate nucleus (LGN) relay cells. When this population of LGN neurons is activated, because their axons contact multiple cortical neurons \citep{Blasdel1983-wp}, the result is a pattern of activity in the cortex. Seen from the perspective of cortical neurons' response properties, cortical neurons' receptive field size and scatter means that many neurons' receptive fields overlap a small spot of light \citep{Bonin2011-ee,Smith2010-pe,Hetherington1999-hb}. This suggests the natural input to cortical networks is not input to a single neuron or just a few neurons, but a population pattern of input to many cortical neurons.

Thus, patterns of input across neurons are the way natural inputs often arrive. Patterned inputs explain how robust recurrent network effects can arise in the cortex, in spite of weak coupling between many cells. Many cortical recurrent synapses depolarize neurons only a small fraction of the distance between rest and threshold \citep{Holmgren2003-vx}, and the probability of connection between any two cells is well below one \citep{Ko2013-ow}. This is consistent with the findings, from both electrical and optogenetic stimulation experiments, that individual neuron activation \textit{in vivo} produces detectable --- but small --- effects on both the local network and on behavior \citep{Chettih2019-gf,Dalgleish2020-lk,Houweling2008-lx}.

However, because natural input changes the activity of not just one, but many neurons, patterned or population natural input can produce postsynaptic summation, allowing many weak local recurrent inputs to sum and produce large responses in target cells. Thus, while input to one cell will have small or weak recurrent effects on other cells, population input can engage the local recurrent network \citep{Dalgleish2020-lk} and can produce large effects. A pattern of long-range input can create a pattern of recurrent input that interacts and combines with the long-range inputs. The local recurrent effects modify responses to the input activity pattern, so that the pattern of activation in the cortex differs from the pattern of input that arrives \citep{O-Rawe2023-ya,Sanzeni2022-ko,Mulholland2024-fu}. This is the foundation of the active filtering computation: the recurrent network transforms patterns of input to different patterns of output firing.

The active filtering computation seems likely to be a common computation in many sensory areas (see section: ``Active filtering across sensory areas''). Beyond sensory cortex, it may also apply broadly across other cortical areas, as areas that mediate motor and cognitive tasks may also need to transform inputs to select some subset of preferred input.

One effect of active filtering is to amplify input patterns associated with natural sensation (section ``Evidence in sensory cortex for pattern amplification''), and suppress or filter out other patterns, such as the ongoing variable firing that occurs in spontaneous activity \citep{Ringach2009-oh,Amit1997-uj,Burns1976-jo,Kerr2005-gh}.

It is likely that the transformation of input seen in active filtering performs not just amplification but other operations as well. First, the recurrent network can shape representational geometry. The different representations of sensory input in different cortical regions are created in part by feedforward and feedback connectivity, and are also likely shaped by the recurrent network. In addition, both sparsification \citep{Deveau2024-ln,Vinje2000-ue}, and straightening \citep{Henaff2021-ny}, seem likely to occur with active filtering ---~producing effects that accompany the pattern amplification ---~and both can improve coding and therefore behavioral function. Also, other examples of network-level computations likely due to recurrent connectivity have been identified, including computation by dynamics (decision by dynamics) \citep{Mante2013-tn} and normalization \citep{Carandini2011-bx}. Others include pattern separation and pattern completion, as seen in hippocampal regions thought to be related to memory processing \citep{Knierim2016-nt,Aimone2011-um}. (See section: ``Relation to other neural computations.'') Our understanding of brain function is likely to be advanced significantly in future as we classify these network-level computations and determine the circuit mechanisms that create them.

In principle, active filtering could be created by a purely feedforward network, and in fact deep networks for vision generate different representations by transforming input sequentially purely with feedforward synaptic connectivity. However, in biological brains, feedforward pathways are information bottlenecks \citep{Semedo2019-ez} --- there are fewer feedforward synapses than recurrent synapses. This, combined with experimental evidence that patterned input generates local recurrent effects \citep{Dalgleish2020-lk,Sanzeni2023-mw,O-Rawe2023-ya,Marshel2019-fn} , suggests that the recurrent network is a major player in active filtering.

The field of neuroscience is currently at an early stage in understanding network-level or population-level neural computations despite their likely importance. Largely, this is because interconnected networks like those in the cortex are difficult to analyze without experimental means of perturbing neurons. The strongest data to support the active filtering recurrent computation have come from stimulation or perturbation methods, and the coming years promise to bring to bear new data from stimulation experiments from many different brain areas and layers to understand brain network computations (See section, ``New tools\dots{}'').

The study of brain network computations is likely to have a major impact on human health by shedding light on how genes and genetic changes affect behavior. Studies that manipulate genes, or study genetic variation, have not produced many strong links between single neuron-related genes and behavioral phenotypes or pathologies \citep{Torkamani2018-pd,Martin2019-ia}. Instead, it has been observed that a large number of genetic changes each have small influence on behavior or on pathology (as quantified by polygenic scores \citep{Hyman2018-dx}). In this circumstance, how can we link genetic variation and behavior? Put another way, how can we hope to treat diseases using genetics if many hundreds of genes each convey only a small change in risk? Understanding principles of neural computation is a promising path to bridge this gap between genes and behavior. Genetic changes affect circuits and synapses, and circuits and synapses control computations. In turn, these neural computations create behavior: memory, sensation, action, and cognition. Therefore, understanding how genetic changes affect computations is likely to shed light on the key brain changes that create pathology. For this reason, current and future studies of neural computations, computations like active filtering, promise to significantly advance our knowledge of brain function and point towards future treatments. The future for work on neural computations should be exciting.

\section{Neural computations and sensory cortex: what is the sensory recurrent network doing?}

The theories that we have for recurrent network function are often generative. In prefrontal cortex, it is thought that delay activity that supports working memory is generated by recurrent connectivity: the subject perceives a stimulus, the stimulus disappears, and then the network generates ongoing or sustained activity associated with the memory \citep{Funahashi1993-ri,Fuster1991-gc,Narayanan2009-pd,Sreenivasan2019-xv}. Similarly, motor cortex is thought to generate slow dynamic responses that are associated with different complex movements \citep{Svoboda2018-oq,Churchland2012-tx,Russo2020-xx,Hennequin2014-hr,Sauerbrei2020-yr}.

These generative cortical responses can be high-dimensional. Even if the sustained activity or slow dynamics associated with one memory or movement is low-dimensional \citep{Sussillo2013-ez,Ganguli2008-ak,Marton2022-fc}, responses to many stimuli or motions can make up a higher-dimensional space. The large number of local recurrent connections in each cortical area, which has many dimensions or parameters, is therefore well-placed to shape generative dynamics.

In sensory cortex, however, most activity is not generative, but time-locked to sensory inputs. A sensory stimulus arrives, neurons respond. And when the stimulus ends, the neural response soon disappears. Without generative dynamics in sensory cortex, what kind of computations does the recurrent network perform?

Active filtering, boosting relevant patterns of input and suppressing others, is a computation that can support complex, high-dimensional transformations in sensory cortex, even without generative or sustained responses. In sensory systems, inputs arrive to a cortical network when animals or humans sense stimuli ---~for example, when they see a face, touch an object, or hear a voice. Those inputs can be quite complex. Take a subject navigating through the natural world, or viewing a natural movie. Each individual frame of the movie contains many visual features, and this produces a complex pattern of inputs from the periphery that arrive to many different visual cortical neurons. As the movie unfolds, the visual experience changes and the set of inputs to the cortex changes as well.

Thus, during natural visual experience, the visual cortex experiences a changing pattern of inputs that arrive to populations of neurons. This is true for other sensory systems as well. For example, during music or a spoken lecture, the auditory cortex also experiences changing patterns of inputs to its neurons.

\begin{figure}[!ht]\centering
    \includegraphics[width=0.8\textwidth]{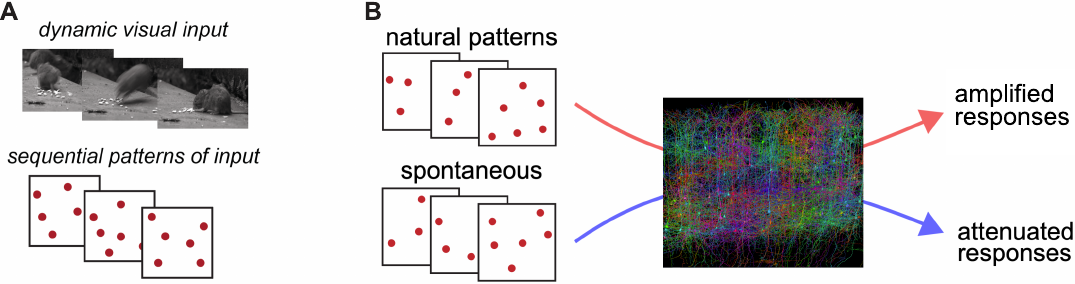}
    \caption{Active filtering: conceptual schematic.
    \textbf{A}, Natural sensory input, in this case visual input, arrives to the cortex as changing patterns of synaptic input across neurons.
    \textbf{B}, These patterns or sequences of patterns arrive to the cortex and are modified by the local recurrent network. Some patterns, those associated with natural sensation (by learning or development or both), are amplified. Other patterns, such as those that might arise from spontaneous activity or other forms of variability or noise, are relatively attenuated.}\label{Fig:2}
\end{figure}

Active filtering operates on those input patterns and changes them, amplifying some patterns or sequences of patterns and suppressing others. Accumulating evidence from several different sensory systems \citep{Peron2020-xz,Marshel2019-fn,Oldenburg2022-nw,Deveau2024-ln} suggests that it is natural patterns of input, the input patterns that reflect natural sensory experiences and natural sensory statistics, that are selectively amplified (Fig.~\ref{Fig:2}).

Therefore, active filtering is a complex, high-dimensional computation, but \textit{not} a generative computation. Inputs arrive to a population of neurons, and as they arrive, the recurrent network modifies the inputs to produce a different output pattern of activity. When the inputs end, the cortical responses also end. This is notably different than in other sorts of recurrent dynamics \citep{Seely2016-zy} where dynamics seem to be generated locally, as for example in motor-related cortical dynamics. In motor planning or execution, neural activity seems to evolve through a low-dimensional subspace of activity that is gated by (low-dimensional) input, but generated in the cortical recurrent network \citep{Sauerbrei2020-yr}.

In contrast, sensory active filtering is an input--output transformation where high-dimensional inputs are transformed by the cortical network to high-dimensional output activity patterns, with dynamics that reflect the input dynamics.

\section{Neural computation: mapping inputs to activity patterns}

A neural computation is a transformation from a set of inputs to a set of outputs. For example, in a cortical network, inputs arrive as axons projecting to the cortical area fire. This produces patterns of synaptic release, which create spiking patterns of activity in different neurons in the network. The neural computation is therefore the transformation from input spikes that arrive to patterns of spiking in the network. This can be a very simple transformation. In some circuits, the input pattern could in principle evoke a mirrored pattern of activity in target neurons, perhaps to trigger one subnetwork vs another. Or, the properties of local neurons, their dendrites, and the structure of local circuit connectivity can produce very complex transformations from input patterns to output patterns.

This conception of a computation --- one input pattern yields one output pattern ---~can also include dynamically changing inputs and outputs. In general, neural computations transform a time-varying pattern of inputs into a time-varying pattern of outputs. Networks that process brief or changing input can still be performing computations, generating output patterns that are transformations of the input they receive.

This work uses neural computation in this way: to mean a transformation of a pattern of inputs to a pattern of output spiking in a neural population. One might also call this the network- or population- level input--output function of the network. Active filtering is a process by which a pattern of input arrives to a cortical network and produces a pattern of output firing that is determined by connectivity. Some of this transformation may be due to feedforward connectivity, but accumulating evidence suggests that the local recurrent network plays an important role in this computation.

We can also think about neural computations in phase space, describing the activity of a network and how it changes over time by assigning one dimension to the firing rate of a single neuron. Past work on `neural manifolds' \citep{Jazayeri2017-pg} postulates that lower-dimensional subspaces, with each dimension in the subspace or manifold defined as a combination of single neurons, are good ways to describe the operation of neural networks. In this framework, each output pattern of activity is a point in the phase space. Active filtering reshapes the relationship between input and output in this space. Input patterns specify a point in the phase space: when input patterns arrive, they drive the network to an activity state. This happens quickly, with the network reaching a steady state to a given input in a few tens of milliseconds \citep{O-Rawe2023-ya,Sanzeni2020-ei,Ozeki2009-lu}. The connectivity in the recurrent network changes that mapping, from input patterns to output activity. In other words, a recurrent neural network's computation can be put in terms of neural manifolds: the neural manifold accessed during normal function would have one shape without the recurrent network, but the recurrent connections change the shape of that manifold.

\section{Evidence in sensory cortex for pattern amplification, the building block of active filtering}

The core computation underlying active filtering by sensory cortex is selective amplification of one pattern relative to another. The network in a sensory cortical area can experience many different patterns of activity. In active filtering, some of those input patterns are privileged --- selected --- by amplifying those patterns, and some of the input patterns are relatively suppressed.

Thus, the network operation that underlies active filtering is selective amplification, a network response that is larger for some spatiotemporal patterns of input than for others. The important aspect of active filtering is to separate different patterns of activity by the recurrent network boosting some patterns and not others. The absolute magnitude of the resulting pattern of activity in the network is likely less important --- both `amplified' and `suppressed' responses might be larger than some reference response, or both smaller. The absolute magnitude of amplified and suppressed responses in a network can be scaled up or down together in simple ways \citep{Chance2002-py,Pouille2009-kd}. This can be done without changing information content, if as seems likely, single-neuron variability is not a form of noise that significantly limits behavioral performance \citep{Beck2012-wh}. This scaling would not affect the important part of the active filtering computation, selective amplification ---~the separation of responses of the amplified and suppressed patterns from each other.

A few important observations of selective amplification have arrived within the past few years. Despite the large number of ongoing studies of sensory cortex, experimental observations of amplification are fairly recent, because methods to causally affect specific neurons in the cortex have become recently available.

Patterned stimulation work, using two-photon optogenetics, which can induce patterns of activity in the cortex directly, has provided strong evidence for cortical pattern amplification. Input patterns that arrive on visual cortex neurons with similar orientation tuning are amplified \citep{Marshel2019-fn,Oldenburg2024-dc}, consistent with prior anatomical evidence \citep{Ko2011-mo,Lee2016-pg}. Input patterns that are associated with natural vision are amplified when those inputs are played back onto visual cortex neurons \citep{Deveau2024-ln}. And cell-specific lesion studies have shown that input patterns that arrive on somatosensory cortex neurons with similar whisker tuning are also amplified \citep{Peron2020-xz}.

These observations of amplification are both relatively new and create a different conception of cortical function than had previously existed. Before these types of stimulation experiments, it was difficult to infer input--output transformations directly. The classic studies of Hubel and Wiesel highlight this difficulty. Hubel and Wiesel generated one of the most important steps forward in neural computation to date: they studied the transformation between LGN neuron inputs and cortical simple cell firing, showing that center-surround LGN inputs could sum inside cortical neurons to create edge-detecting simple cells \citep{Hubel1962-yd}. But it was not until Marshel et al. (2019) that it was made clear that many of those oriented-edge cells in the cortex, when activated together, could self-amplify via local recurrent interactions. Before the work of Marshel et al. we knew that about two-thirds of the input to visual cortical neurons came from the recurrent network \citep{Lien2013-en,Chung1998-tk,Li2013-ni}. Perhaps we could have guessed that recurrent interaction was creating specific pattern amplification, but many careful and rigorous manipulations over the years, primarily elucidating excitatory--inhibitory interactions, were also consistent with feedforward connectivity alone, not recurrent amplification, creating these cortical responses \citep{Priebe2012-qs}. And in principle, it could have been that feedforward connectivity alone is sufficient to perform all the neural computation in sensory cortex --- as seen in deep convolutional artificial networks for vision \citep{Krizhevsky2017-ne}, which do not rely on recurrent computation. It is patterned stimulation approaches, which allow study of how excitatory cells affect one another by controlling patterns of input directly, that has now firmly established this amplification of population-level inputs in the cortex.

\section{Relation to other neural computations: from pattern completion to normalization}

Several existing network computations are related to active filtering of natural input. Pattern completion, where a partial pattern is boosted by the network to yield a fuller pattern, is a form of filtering. Pattern completion is closely related to amplification, and in fact, amplifying a pattern of input can be a way to in practice achieve pattern completion. Future work should shed light on whether the cortex is performing pattern completion, with an attractor state at the completed pattern, or in contrast whether there is amplification along a line in phase space without a fixed point \citep{Khona2022-py}, approximating a metastable or continuous attractor as would be expected for a pattern amplification process. Experiments that vary pattern strength to determine whether completed or output responses vary with input strength could reveal whether pattern amplification with purely stable attractor dynamics is occurring.

Another related computation is normalization. Normalization is the observation that co-occurring inputs produce smaller responses in single neurons than would be predicted by the sum of the responses to each input alone. Normalization, like active filtering, is also a kind of input--output transformation produced by network dynamics, but it is likely a separate and orthogonal computation from active filtering --- that is, both computations operate on cortical inputs but the strength of normalization does not dictate the strength of active filtering or vice versa. Normalization describes responses to complex input patterns, allowing prediction of responses to a complex input from responses to simpler component inputs. Meanwhile, active filtering also describes responses to complex inputs, but speaks to the comparative strength of response due to one input compared to other inputs. Active filtering is a process that requires learning the statistics of the sensory world, while the normalization computation can be learned purely from sets of input patterns, so it may be that active filtering is more directly related to the organism's behavior.

Finally, active filtering may be related to representational straightening \citep{Henaff2019-mm,Henaff2021-ny}. Physiological and psychophysical experiments have shown that the visual system transforms representations in early visual areas to produce representations in later areas that are less curved in pixel or neural activity space. It seems likely that the mechanism for straightening is the transformation created during active filtering, which amplify but can also sparsify input patterns, modifying them \citep{Deveau2024-ln}.

\section{New tools are enabling study of brain network computations}

Recurrent networks like those in the brain are difficult to study. One major reason for this is the interconnectedness amongst neurons makes it hard to determine the cause of any observed neural change. When one neuron in the cortex changes its firing rate, recurrent connections mean that other local neurons' firing rates are also affected.

When a sensory stimulus changes, neurons in the cortex change their firing rates. But how do we know whether these firing rate changes arise from long-range inputs external to the network, or arise from local connectivity?

In the past decade, new methods have become available for recording large numbers of neurons as the brain operates under normal, awake conditions: Utah electrode arrays, Neuropixels recording probes, and mesoscopic two-photon imaging, to give just some examples of these approaches.

But even large-scale neural recordings have difficulty determining how networks respond to input. In sensory cortical areas, feedforward inputs arrive from thalamus or cortical areas closer to thalamic input. Beyond feedforward inputs, the pyramidal cells themselves send synapses to other local pyramidal neurons. Add in feedback connections from other brain regions, and cortical pyramidal cells receive a huge number of different inputs that are modulated by a sensory stimulus. A similar situation holds for other kinds of cortical processing. Dynamics in cortical activity involves populations of cortical pyramidal cells that receive inputs from many other neurons, both locally and from distant areas.

Using recording methods alone to infer the effects of all these connections requires inferring how each input affects neurons' activity. In principle, this can be done, but in practice, it is often difficult to do without recording many or all neurons, not just in the cortical area but in input areas as well.

Stimulation methods get around these problems by directly changing input to selected neurons, allowing direct inference of how the change in activity of one neuron, or a population of neurons, affects the network \citep{Jazayeri2017-pg,Pearl2013-ic}. Recording approaches are still needed to monitor neural activity, and experiments that combine stimulation and large-scale recording seem best-placed to study recurrent network dynamics.

A variety of causal or stimulation methods have helped us make progress understanding recurrent networks, including chemogenetics, electrical stimulation, transcranial magnetic stimulation, and focal lesion methods \citep{Peron2020-xz}. One powerful stimulation method stands out: two-photon optogenetics, or two-photon holographic stimulation. Because the normal input to cortical networks is a pattern of input across many neurons, understanding how recurrent networks process their inputs depends on the ability to change the activity of patterns or populations of neurons --- not just one neuron or a handful of neurons. Two-photon holographic optogenetics provides this ability \citep{LaFosse2023-yy,Oron2012-pr,Packer2012-iw}. It uses a spatial light modulator as a programmable beam splitter to target stimulation laser spots to populations of neurons at cellular resolution. Using this approach, patterns of neurons can be stimulated, allowing us to answer causal questions of how recurrent networks are influenced by patterns of input.

Because stimulation or causal methods are essential for understanding interconnected networks, and cortical inputs arrive as patterns, two-photon optogenetics is a tool with particular promise for advancing our understanding of recurrent cortical networks.

\section{The value of theoretical models for understanding interconnected networks}

Along with progress in experimental tools, another reason to be optimistic about the future of understanding of brain recurrent networks are progress on theoretical and mathematical frameworks. In interconnected networks like those in the brain, changes in one part of the network can have unintuitive or paradoxical effects across the network. Experimental approaches, where activity patterns are changed one by one and responses measured will alone make only slow progress. General understanding of network properties will require building models. As in other fields (e.g. economics \citep{Christiano2018-co}) the most productive way forward is likely to be an iterative process between model and data, where experiments are used to study different pieces of the network, the model is updated, predictions result, and these predictions are tested.

Two major types of theoretical work are already informing the study of brain recurrent networks: work in computational neuroscience and in artificial intelligence~(AI) or machine learning. Computational neuroscience has shed light on many aspects of how circuits can affect networks, by developing models of recurrent networks with varying levels of network and single neuron complexity. And the field of AI, by characterizing large scale models \citep[e.g.][]{Rudin2022-mf}, that are made up of simple units only roughly inspired by biology, has begun to add information on the computational principles of networks more generally. Experimentalists will now have their work cut out for them in testing predictions from these theories to refine future models.

In summary, understanding complex interconnected networks require causal stimulation methods, the ability to record and monitor many components of the system, and a close interaction between theory and experiment. Great progress has been made on all three directions in recent years, and all of these pieces now seem to be in place to rapidly advance our knowledge of brain recurrent networks.

\section{Pattern amplification depends on strong recurrent interactions: the cortical balanced state, and inhibition-stabilized network models}

How is pattern amplification created in a recurrent network like that of the cortex? Artificial RNNs are universal function approximators \citep{Schafer2006-rr} that can be trained to perform amplification of some patterns and not others. But what is known about how more biologically relevant neural network models, with neurons that fire spikes and with excitatory and inhibitory populations, might support amplification?

Principles of pattern amplification have emerged from the study of recurrent network models with excitatory and inhibitory populations. If the local recurrent connectivity is strong enough, recurrent weights or connections can be structured to amplify some patterns and not others \citep{Murphy2009-il,Goldman2009-az}. Theoretical study of this amplification mechanism, dubbed ``balanced amplification'' by Murphy and Miller, has also shown that the dynamics of the resulting amplification can be fast --- that is, the network responds quickly as input is applied, and responses dissipate quickly when the input is removed \citep{Goldman2009-az,Murphy2009-il}. These fast responses are what is seen in the cortical network \textit{in vivo} as well: responses to sensory stimuli begin on stimulus onset and end within a few tens of milliseconds after stimulus end. Recurrent amplification of this sort can amplify many different patterns, via different patterns stored in the many network weights. This is consistent with the large number of recurrent excitatory--excitatory synapses in cortical areas, which give a large number of parameters to encode patterns into memory \citep{Hopfield1982-vz,Krotov2016-ei,Little1978-nf,Aljadeff2021-db}.

To allow pattern amplification through balanced amplification, the recurrent network should be operating in the balanced state, with strong overall excitatory connectivity. The balanced state \citep{Van_Vreeswijk1996-sl,Brunel2000-em,Wolf2014-ex,Destexhe2003-bc} is a regime of recurrent network operation where neurons receive large amounts of excitatory input that is balanced by large amounts of inhibitory input. This observation is widely supported by experimental data \citep{Ahmadian2021-bo}.

The balanced state specifically refers to excitatory--inhibitory interactions, and is built on the observation that neurons in the cortex receive large amounts of excitation and inhibition that track one another \citep{Okun2008-vq}. The large excitatory and inhibitory input conductances balance, or cancel, to leave a membrane potential at the soma that has a small mean, but significant fluctuations. Because neurons' membrane potential fluctuates, it is at times near threshold which results in irregular firing \citep{Brunel2000-em,Vogels2009-jm,Sanzeni2022-xk,Alexandersen2024-ik}. (As a side note, this property of cortical neurons in the balanced state ---~irregular threshold crossings --- explains the approximately-Poisson firing seen in cortical neurons. Under assumptions about autocorrelation decay that are present in the cortex \citep{Bair2001-gz}, the spike threshold crossings form a ``baby Bernoulli'' process \citep{Gallager2012-wk} that approximates a Poisson point process.)

When the membrane potential of a cortical neuron fluctuates, even weak inputs, as seen for excitatory cortical inputs \citep{Bruno2006-qn}, can change a cell's firing rate. If neurons were not balanced, if a cell was receiving strong excitation or inhibition that was unbalanced, it would either fire at a very high rate, or be deeply hyperpolarized and not responsive to input. Thus, one important consequence of the balanced state is that neurons can convey information about even weak inputs in their firing rates.

\begin{figure*}[!p]\centering
    \includegraphics[width=0.65\textwidth]{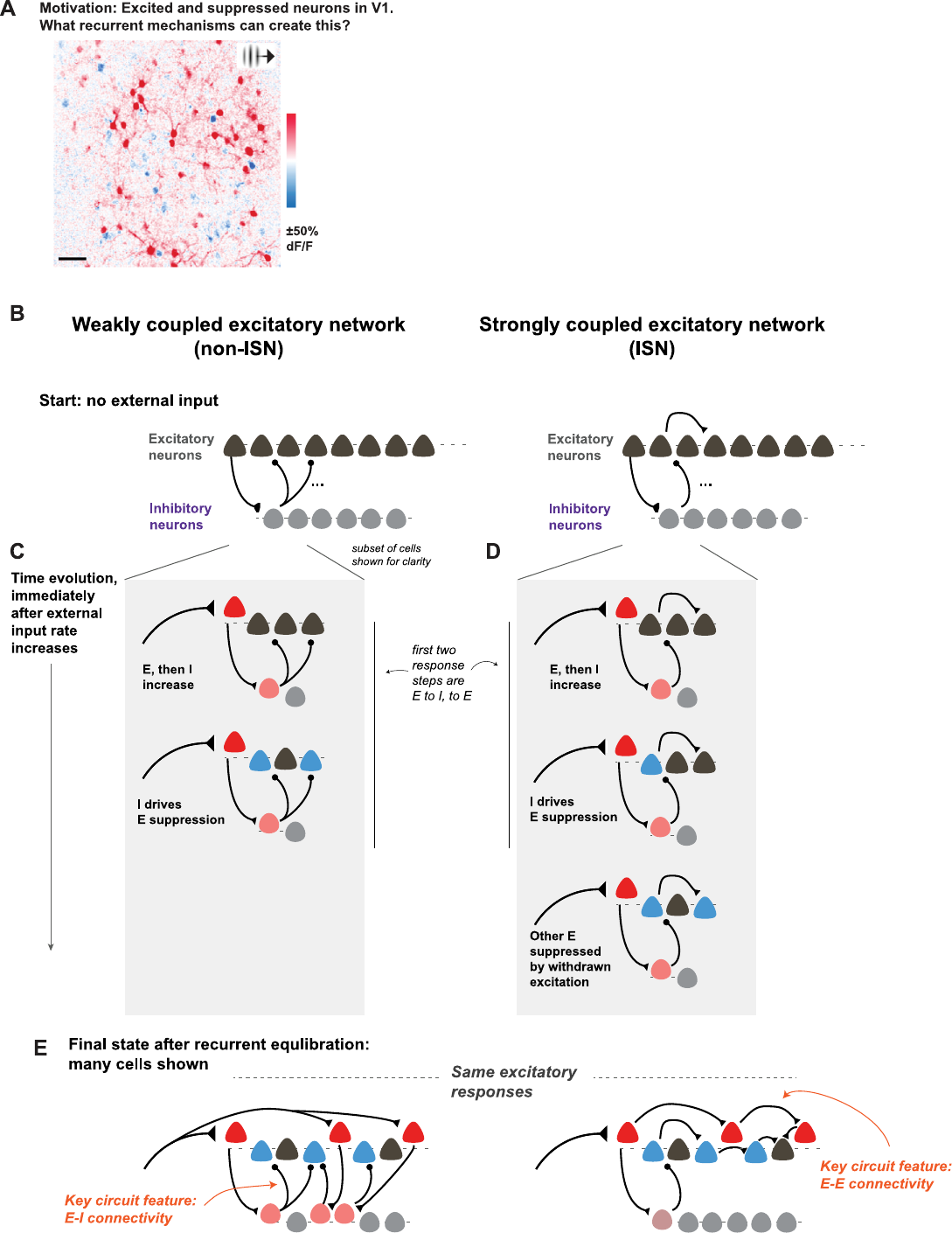}
    \caption{Inhibition stabilization and strong coupling: the strong cortical recurrent excitatory connectivity can control responses to input.
        \textbf{A}, Example response in mouse L2/3 visual cortex imaged with a two-photon microscope. The subject is awake and viewing a small drifting grating patch \citep[methods described in][]{LaFosse2023-yy,O-Rawe2023-ya}. Some neurons are excited and some are suppressed.
        \textbf{B}, If this suppression is occurring via local recurrent mechanisms, what are those mechanisms? Two possibilities can be distinguished: the weakly (excitatory-excitatory) coupled mechanism (left) and the strongly-coupled mechanism (right). The difference is whether local excitatory-excitatory connections play an important role or not. Schematics: cortical networks composed of excitatory and inhibitory cells. Connections between a few E and I cells are shown to explain initial responses to input.
        \textbf{C}, Schematic of the first few milliseconds when input arrives at the weakly-coupled excitatory network. Input excites an excitatory cell, which excites an inhibitory cell, which suppresses some other excitatory cells.
        \textbf{D}, Same, but for the strongly-coupled network. Input excites an excitatory cell, which excites an inhibitory cell, which suppresses an excitatory cell. Here, however, the strong excitatory connections result in excitation being withdrawn from other local excitatory cells. The network settles into a new steady state that depends both on the input pattern and on the recurrent inputs. Here input is shown arriving at excitatory cells only.
        \textbf{E}, Steady-state pattern of responses, approximately \qty{30}{\ms} after the first input turns on \citep{O-Rawe2023-ya,Sanzeni2020-ei}. The two circuit configurations, one with strong excitatory connections and one with weak, can produce the same pattern of excitatory cell firing rates via different key circuit mechanisms: specific I--E connectivity (left) and specific E--E connectivity (right).
        \textbf{Notes:} Feedforward inputs arrive at inhibitory neurons as well, omitted here without changing the core phenomenon. Also, here the inhibitory cells are not distinguished by subtype; inhibitory neurons shown may comprise several subtypes without changing the core argument. Last, only a small number of excitatory and inhibitory synapses are shown. While individual cortical synapses are often weak, patterns of excitatory cell activation can produce the effects shown in E as inputs from many local E cells sum on individual neurons.
    }\label{Fig:3}
\end{figure*}

The balanced state is connected to the idea of strong excitatory connectivity through a framework called the inhibition stabilized network (ISN) model \citep{Sadeh2021-mg,Tsodyks1997-pm} (Fig.~\ref{Fig:3}). In the ISN regime, aggregate local excitatory influences are strong enough so that the excitatory network would be unstable. That is, the cortical network would explode with unrestrained activity, if inhibition did not react to compensate for changes in excitation. The balanced state and ISN regime of cortical operation are closely related --- the balanced state largely describes the excitatory--inhibitory interactions, while the ISN principle is that the excitatory--excitatory recurrent network is unstable. But if the network is an ISN, it must be (as the name implies) stabilized by inhibition. Thus, while networks can be in excitatory--inhibitory balance but not be an ISN, the ISN state generally implies a balanced state.

Past work provided evidence for the balanced state but had provided somewhat contradictory results on key experimental predictions of the ISN model \citep{Atallah2012-lx,Moore2018-gn,Kato2017-yl}, calling into question which, if any, cortical areas might operate in the ISN state. One possibility was that because individual cortical excitatory synapses are relatively weak, their total effect on the network was also weak despite the number of excitatory--excitatory local synapses. However, a set of recent studies \citep{Sanzeni2020-ei,Sanzeni2022-ko,ORawe2022-pc,Kato2017-yl,Li2019-de,Sadeh2021-mg} \citep[but see][]{Mahrach2020-jk} have provided evidence that cortical recurrent connectivity is strong in several cortical areas, confirming pioneering work in visual cortex \citep{Ozeki2009-lu}. The studies also find evidence for strong excitatory recurrent interactions across species: visual cortex of both the rodent and non-human primate. Together, these studies support the idea that the ISN state, and thus strong local excitatory--excitatory recurrence, is a common feature of many if not all cortical areas.

Taken together, these experimental and theoretical results provide a foundation that can support pattern amplification. Cortical recurrent connectivity is strong, implying that local recurrent connections can in principle change input--output transformations. Specific stimulation and specific cell lesion work \citep{Sanzeni2022-ko,ORawe2022-pc,Peron2020-xz,Marshel2019-fn} demonstrate that network-level input--output transformations can indeed be changed by recurrent interactions, providing a foundation for pattern amplification. And the balanced amplification model shows how pattern amplification can occur in networks like the cortex.

What results is a conceptual framework for sensory cortex that builds on the balanced state framework. The role of excitatory--inhibitory balance in the cortex is in part to allow the excitatory--excitatory recurrent network to be self-amplifying, and that local excitatory feedback is used to support computations like pattern amplification ---~boosting some patterns of input and not others.

\section{Active filtering across sensory modalities}

In rodent visual and somatosensory cortex, evidence from stimulation studies shows that naturally-occuring patterns of input are amplified \citep{Marshel2019-fn,Peron2020-xz,Deveau2024-ln,O-Rawe2023-ya}. How general is this across sensory areas?

Work in auditory cortex suggests that auditory circuits also perform active filtering. In the rodent, silencing experiments show that excitatory recurrent inputs to auditory cortical neurons are sharply frequency-tuned, suggesting some recurrent amplification of sounds based on their frequency \citep{Liu2007-wv}. Other work combining inhibitory silencing and theoretical models also suggest recurrent amplification in auditory regions \citep{Aponte2021-ce}.

Moreover, when active filtering operates in the time domain, this yields neural responses to dynamic stimuli that are modulated based on stimulus history. If this is true, we might expect a signature to arise as some sort of sustained response in neural activity after a stimulus turns off, even if this is brief or the mean response across neurons is weak.

First, supporting this idea, an offset response in primary auditory cortex (A1) has been used to argue for recurrent amplification \citep{Bondanelli2021-im}.

Second, it is widely reported that moderate- or stronger-intensity stimulation of many cortical areas produces offset effects that last for a few hundred milliseconds after the stimulus ends \citep[e.g.][]{Butovas2003-wf,Sombeck2022-pc}. That is, when a population of excitatory cells in the cortex is driven to a higher firing rate, some neurons are suppressed after the stimulation turns off. This suppression may be a signature of active filtering that extends over time, as seen in visual cortex as well \citep{Deveau2024-ln}. Although some neurons are excited and some suppressed by recurrent interactions, the average response is suppressive, and lasts for a few hundred milliseconds, allowing earlier inputs to interact with later-arriving inputs. The fact that this offset suppression response is seen in many cortical areas may suggest that active filtering across time is also occurring in many cortical areas.

From an organismal, or ethological, point of view, it makes sense that many sensory areas may perform the neural computation of active filtering. Many sensory modalities receive high-dimensional signals that change in time --- music or vocalizations for audition, dynamic movies or complex motion for vision, the complex interaction of an object with a moving set of whiskers in the somatosensory system. Each of these functions would seem to benefit from selecting naturally-occuring patterns of input and amplifying them relative to random patterns.

Even more broadly, artificial recurrent networks (RNNs) in machine learning systems are often used to process temporal signals, and often used to produce, generate, or complete sequences (Fig.~\ref{Fig:3}). This type of generative sequential processing may happen in the motor system or the frontal cortex, where complex dynamics associated with behavior have been observed. In sensory cortex, such sequence processing may be happening through active filtering: modifying dynamic input without generative dynamics. Taking these observations together, it seems possible that processing sequences of patterns of input is a major cortical function --- perhaps \textit{the} major cortical network function. The convergence of network theory and experiments, especially stimulation experiments, that is happening now in systems neuroscience seems likely to shed more light on the general importance of sequence processing in many cortical regions.

\section{Recurrent network effects and dendritic or single-cell nonlinearities can co-exist, and may both contribute to active filtering}

Active filtering is a process by which certain patterns of input are privileged over others, by boosting, pattern completion, or amplification by the cortex. Up to this point, I have described this effect by assuming that the amplification process is created by the recurrent connectivity. That is, one can think of the local recurrent connections as being described by a synaptic weight matrix, and then the choice of weights --- subject to additional variation over time via effects like short-term plasticity and spike-frequency adaptation --- is what creates the transformation from patterned input to population or patterned spiking output.

But an additional prominent source of nonlinearity in neuronal networks arises from single-cell \ effects. Cortical excitatory neurons have dendrites that contain many active processes that can produce nonlinear summation of inputs \citep{Stuart2016-oc}. Dendritic nonlinearities may also be a mechanism for active filtering; a way to boost certain inputs. For example, imagine a subnetwork of neurons in a cortical area, and consider just one single cell that the input subnetwork synapses on. It could easily be the case that the synapses on the postsynaptic cell are arranged to produce nonlinear amplification of input from that particular subnetwork. Then dendritic nonlinearities could create a large amplified effect at the soma that produces different and larger spiking than equivalently-sized inputs from a different subnetwork.

Both of these mechanisms --- amplification via the network's synaptic weights, and nonlinear dendritic amplification ---~could be a substrate for active filtering and pattern amplification. They may both operate at the same time.

That said, there are a few observations that may help point the way forward, as future experimental manipulations aim to determine the relative importance of the two mechanisms for active filtering. First, dendritic nonlinearities are a sufficient but not necessary mechanism. Artificial recurrent neural networks can perform filtering via amplification without dendritic nonlinearities by merely adjusting synaptic weights. There is a long history of dynamic neural computations created via weight matrices, from early connectionist models \citep{Minsky1969-bb} to the Hopfield network \citep{Hopfield1982-vz} to more recent discoveries in the hippocampus \citep{Lee2015-rl}. Beyond simple artificial networks with binary or rate-based neurons, more realistic networks that respect Dale's law (i.e., have excitatory and inhibitory model neurons) have also been shown to perform amplification (section ``Pattern amplification depends\dots{}''). So dendritic nonlinearities can selectively amplify input patterns, but so can nonlinearities from synaptic connectivity patterns even if dendrites sum input linearly.

One factor that may tip in favor of the pattern of synaptic connections being the primary mediator of active filtering is the dimensionality involved. The large number of recurrent synapses in a cortical area provides a very high-dimensional space to encode many different input patterns, and create the high-dimensional pattern of responses that exists in sensory cortex \citep{Seely2016-zy,Stringer2019-hz}. By contrast, dendritic nonlinearities can in principle produce powerful amplification effects through voltage-dependent processes and dendritic spikes, but the dimensionality or number of parameters available in nonlinear dendrites may be smaller. This is because there is only one dendritic tree per cell. Despite the many branches and stretches of dendrites between spines, if individual synapses can be plastic or adjusted independently of other synapses, in principle this allows the dimensionality of the weight-based mechanism to be larger as it varies with the square of the number of cells.

Given the complexity of the brain, and the tendency for evolution to adopt many or all mechanisms that are available to solve a given neural computational problem, we cannot rule out other kinds of mechanisms that have not yet been well-delineated. Future work will shed more light on this question, including the relationship between these two potential mechanisms for active filtering, and in fact may find that both operate together to create the active filtering effect.

\begin{figure}[!ht]\centering
    \includegraphics[width=0.65\textwidth]{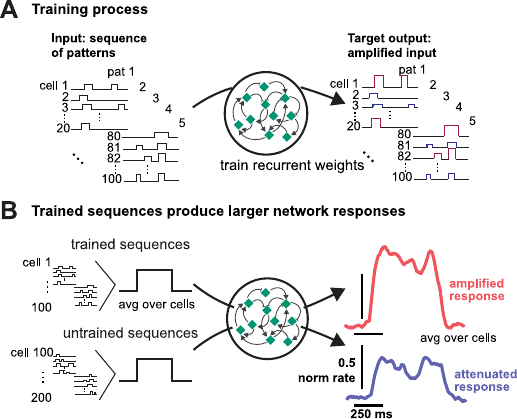}
    \caption{Artificial RNNs can demonstrate the mechanism of active filtering.
        \textbf{A}, An RNN can be trained to filter inputs. Left, a pattern of inputs, varying over neurons and over time, chosen at random. Right, the target output in training: some inputs are amplified. (The network can also be trained to sparsify patterns, as shown here.)
        \textbf{B}, After training, when the network is tested with both trained and untrained sequences, as expected it amplifies the trained sequences and relatively attenuates the other trained sequences. It also shows the fast onset and offset of response, tracking the timecourse of input without generating sustained or ongoing responses, that is characteristic of active filtering. Modified from \citep{Deveau2024-ln}.}\label{Fig:4}
\end{figure}

\section{Whither the engram? Active filtering is one form of engram, a kind of memory}

Active filtering is a predictive process. In active filtering, the way patterns of input are modified depends on learning. This learning may arise from nature or nurture or both: learning from past experience, or learning encoded in genetically-specified circuit structure acquired over evolution. Because active filtering amplifies patterns of input associated with natural sensory input, which patterns are amplified, and which are not, says something about the natural world.

This means that active filtering can also be seen as a form of memory. The recurrent network of the cortex learns about the statistics of the natural world \citep{Berkes2011-ot,Deveau2024-ln}. It uses that information to amplify a subset of input patterns. The amplification process encodes information about the past ---~that is, the identity of the amplified patterns is stored in the network as a memory.

The memory aspect of active filtering is connected to the concept of an ``engram'' as it is often used in neuroscience. The word engram describes the neural basis of a memory \citep{Josselyn2015-so,Josselyn2020-wr}. But is all active filtering a kind of engram, and are all kinds of engrams the same as active filtering? To answer that, we can look back at the definitions of ``engram.''

In 1912 Richard Semon, who coined the term engram, described it in this way:
\begin{quote}
    ``the enduring though primarily latent modification in the [brain] produced by a stimulus, I have called an \textit{engram}. [In this work] we deal with influences that awaken the engram out of its latent state into one of manifested activity. Our study reveals the laws regulating the associations between the latent and the revived [\dots] engrams.''
\end{quote}

There are two definitions of an engram embedded in this text. To cast it in modern terminology, the first definition is the change in the brain that encodes a memory. An example would be synaptic changes. The second, distinct, definition is not the synaptic or other changes when a memory is formed, but instead the activity pattern that results from reviving or invoking that memory.

The first definition fits well with the memory process that is involved in active filtering, where cortical recurrent networks, through experience or development, learn things about the natural world. The filtering relies on an enduring and latent modification in the network. The latent modification is not seen directly in neural activity. But when the organism experiences that stimulus again, the input pattern associated with the stimulus is amplified, creating a pattern of activity in the cortex associated with that stimulus.

While the first definition of engram is consistent with active filtering, the second definition ---~the activity pattern across neurons associated with a memory --- is not. Active filtering is primarily a neural computation that results when a stimulus is present. This is opposed to the kind of reactivation seen in hippocampal engrams \citep{Tonegawa2015-kh,Ramirez2013-hb}, or delay activity in frontal regions which is sustained when a stimulus disappears. Therefore, active filtering does not reflect this second definition --- the reactivation, or activity-pattern definition --- because active filtering does not create activity patterns associated with experiences or stimuli in the past. What active filtering does do, however, is change responses to \textit{current }experiences or stimuli \textit{based }on the past, and the information or memory that allows those changes is indeed an engram in the first sense above, the ``latent brain modification'' sense.

The way the recurrent cortical network encodes the features of the natural world is therefore, in one sense, an engram. There are latent synaptic or cellular changes that allow the recurrent network to privilege some patterns but not others.

Neuroscience and psychology have long classified memory into different types: episodic, spatial, sensory or perceptual, etc \citep{Tulving2004-zq,Alan_Baddeley2013-xc,Eichenbaum2017-hr,Eichenbaum2011-fg,Squire1993-fw}. Many forms of explicit memory appear to be dependent for formation on hippocampal areas, with cortical areas thought to be a final site of storage of long-term memories \citep{Frankland2005-fx,Nadel1997-gx}. However, the focus for a cortical role of memory has often been frontal regions like prefrontal and cingulate cortex \citep{Maviel2004-ka,Tanaka2014-sn}. Neocortical sensory areas have not often been thought of as locations for memory storage. The active filtering computation I describe here may implicate the cortical excitatory--excitatory recurrent network in this type of memory: sensory memories that encode the structure of the natural sensory world. When we remember a face or sing along with a song we know, the recurrent network of the cortex may well be a principal location where those types of sensory memories are stored.

\section{Conclusion}

Recent progress in systems and circuit neuroscience using stimulation methods have provided evidence for active filtering. The local, recurrent synapses in the cortex can encode information about natural sensory input and allow patterns associated with natural sensation to be privileged via amplification, by modifying the input--output function of the network. This is a predictive function but not a predictive code. It is also a form of memory.

What might we expect from future work?

Active filtering is a form of network computation. It is a statement about the way patterns of input are changed to output spiking. It seems likely that major steps forward in understanding brain function will come from understanding network computations. It will be important to enumerate other forms of network computations, determine how they are influenced by brain state, neuromodulators, and input from other brain areas like those involved in emotion and motivation. It will also be important to understand which computations are performed by different brain areas and how simultaneous computations in different brain areas work together or, alternatively, operate separately. That is, during a sensorimotor behavior, do the sensory cortical areas perform separate computations from motor areas, or are they intermingled in fundamental ways? Similar questions can be asked for subnetworks within cortical areas: are they interdependent or performing separate computations?

It is a promising time for the study of these questions. The field of systems neuroscience has developed both experimental tools and theoretical frameworks that now allow study of network computations in functioning brains.

Developing such a classification, or vocabulary, of different types of network computations seems likely to open up new links from brain structure to brain function. The discovery of the double helix structure of DNA was also such a structure-function link. It seems quite possible that the coming years in systems neuroscience will be as exciting for the understanding of how brains operate as the double helix discovery was for the understanding of how cells operate.
\noclub[4]

\bibliography{references}

\end{document}